# Coulomb interaction on pion production in Au+Au collisions at relativistic energies

Catalin Ristea[1,2], Oana Ristea[1], Alexandru Jipa[1]

[1]University of Bucharest, Faculty of Physics, Romania
[2]Institute of Space Science, Magurele, Romania

Email: oana@brahms.fizica.unibuc.ro

**Abstract**: Coulomb effects on charged pion transverse momentum spectra produced in Au+Au collisions at RHIC-BES energies are investigated. From these spectra the $\pi^-/\pi^+$ ratios as a function of transverse momentum are obtained and used to extract the "Coulomb kick", $p_c$ (a momentum change due to Coulomb interaction) and initial pion ratio for three different collision energies and various centrality classes. The Coulomb kick shows a decrease with the increase of beam energy and a clear centrality dependence, with larger values for the most central collisions. The results are connected with the kinetic freeze-out dynamics and discussed.

## 1. Introduction

One of the main goals of the heavy ion collisions at relativistic energies is to study the nuclear matter at extreme energy densities and temperatures. The experimental measurements of the hadron production and transverse momentum distributions provide information about the complex dynamics of the collision and the different phases through which the collision evolves. The system formed in the collision continuously expands and cools until the kinetic (thermal) freeze-out stage, when the produced hadrons no longer strongly interact with each other. Before freeze-out, the dynamics of the system is dominated by strong interactions while the long-range Coulomb interactions become important only after kinetic freeze-out.

In heavy ion collisions at SIS [1,2], AGS [3-5] and SPS [6-8] energies, the transverse momentum spectra of positively and negatively charged pions show a difference at low $p_T$ and these results were interpreted as due to the Coulomb final state interaction between the charged pions and the positive net-charge of the particle source. The net-charge density depends on the degree of the baryon stopping produced in the collision. The collision mechanism is changing with energy from strong baryon stopping at AGS energies, to partial transparency at SPS and RHIC energies and transparent collisions at LHC energies [5,9-13]. At low energies, the colliding nuclei are fully stopped in the collision and the produced fireball expands very slowly. The produced charged particles are moving in a Coulomb field produced by the positive net-charge from the reaction partners. The charged pions, as the most abundantly produced particles and the lightest particles, are influenced strongest by the Coulomb field and therefore, they are accelerated or decelerated and their final (detected) energy is changed. At higher beam energies the collisions have a larger degree of transparency, the net-proton density is reduced and the charged pions are less influenced by the Coulomb field.

The Au+Au collisions at different energies in the beam energy scan program (BES) at RHIC are used to explore the QCD phase diagram and to investigate the properties of nuclear matter characterized by different temperatures and baryon chemical potentials. Based on the BES experimental data, the Coulomb interaction can be studied in a complementary energy domain to AGS and SPS energies.

In this paper, we present an analysis of the Coulomb effects on pion particle production in Au+Au collisions at RHIC-BES energies based on an analytic model developed in Refs [14,15]. The model considers the longitudinal Bjorken expansion of the fireball and assumes that on average, a charged pion will receive a momentum change due to Coulomb interaction or "Coulomb kick", $p_c$

$$p_c \equiv |p_T - p_{T,0}| \cong 2e^2 \frac{dN^{ch}}{dy} \frac{1}{R_f} \quad (1)$$

where $p_{T,0}$ is the transverse momentum at freeze-out, $p_T$ is the final transverse momentum, $dN^{ch}/dy$ is the net-charge distribution and $R_f$ is the kinetic freeze-out radius. The charged pion ratio is

$$\frac{\pi^-}{\pi^+} = \langle \frac{\pi^-}{\pi^+} \rangle \frac{p_T + p_c}{p_T - p_c} exp\left(\frac{m_T^- - m_T^+}{T}\right) \quad (2)$$

where $m_T^\pm = \sqrt{m^2 + (p_T \pm p_c)^2}$, $\langle \frac{\pi^-}{\pi^+} \rangle$ is the total pion ratio and T is the kinetic freeze-out temperature (the temperature at the moment when the hadrons decouple from the system).

## 2. Results

In order to extract information related to the Coulomb interaction between the produced particles we used the charged pion transverse momentum spectra produced in Au+Au collisions at $\sqrt{s_{NN}} = 7.7$, 11.5 and 19.6 GeV and detected with STAR experiment [16]. The experimental data are from Ref [17].

Based on the STAR published charged pion transverse momentum spectra we obtained the $\pi^-/\pi^+$ ratios as a function of transverse momentum for the collision centrality classes of 0–5%, 5–10%, 10–20%, 20-30%, 30–40%, 40– 50%, 50–60%, 60–70%, and 70–80%. The negative to positive pion ratios are shown in Figures 1-3. The $\pi^-/\pi^+$ ratio is close to unity for all studied energies at higher $p_T$. However, at low $p_T$ values an increase of the pion ratio above unity is observed for more central collisions. As the beam energy increases, there is still an asymmetry between π− and π+ production as shown by the low-$p_T$ enhancement in ratios, but these effects are much less significant.

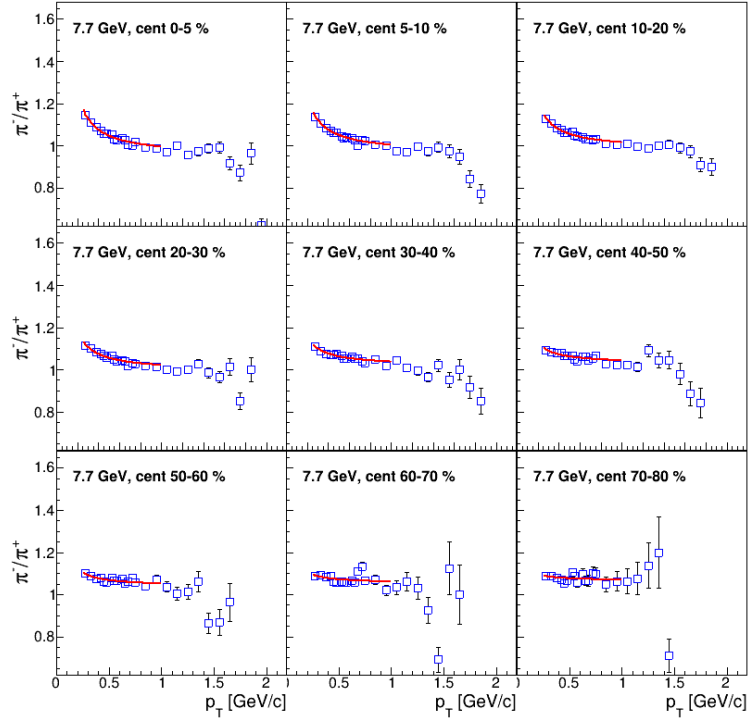

Figure 1: The $\pi^-/\pi^+$ ratios produced in Au+Au collisions at 7.7 GeV as a function of transverse momentum. Experimental data are from [17]. The red lines are the fits with Eq. 2.

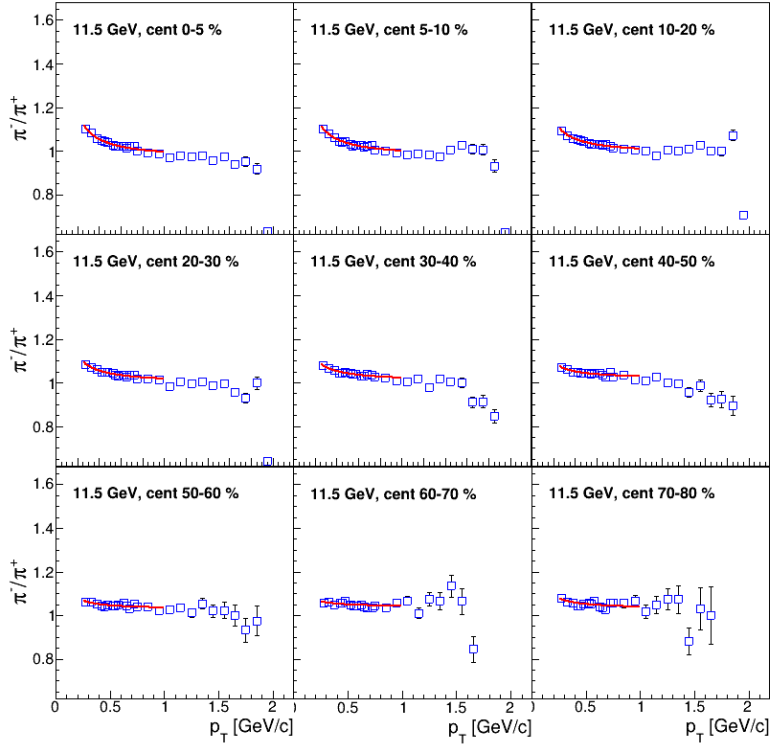

Figure 2: The $\pi^-/\pi^+$ ratios produced in Au+Au collisions at 11.5 GeV as a function of transverse momentum. Experimental data are from [17]. The red lines are the fits to the ratio data using Eq. 2.

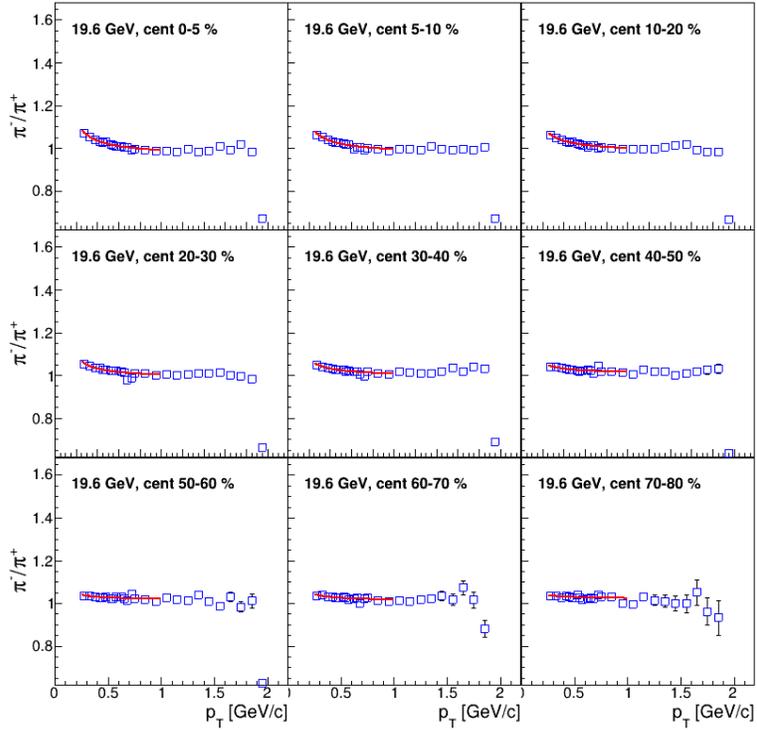

Figure 3: The $\pi^-/\pi^+$ ratios produced in Au+Au collisions at 19.6 GeV as a function of transverse momentum. Experimental data are from [17]. The red lines are the fits to the ratio data using Eq. 2.

The pion ratios were fitted with Eq. 2 to obtain information related to Coulomb effects on pion production. The fits are shown as solid lines in Figs 1-3. The fit range was chosen $0.25 < p_T < 1$ GeV/c, because in the low $p_T$ region the Coulomb interaction has the larger influence on the pion spectra and the ratios. The two free fit parameters are the Coulomb kick, $p_c$, and the total pion ratio, $\langle \pi^-/\pi^+ \rangle$. The kinetic freeze-out temperatures for each beam energy and collision centrality were taken from Ref. [17] and are fixed parameters in this analysis. The fits describe well the charged pion ratios at all energies and collision centralities.

For higher BES beam energies ($\sqrt{s_{NN}} = 27$ GeV and 39 GeV), the STAR experimental data show no significant asymmetry between positive and negative pions at low $p_T$ and it is not possible to obtain information related to Coulomb interaction from those data.

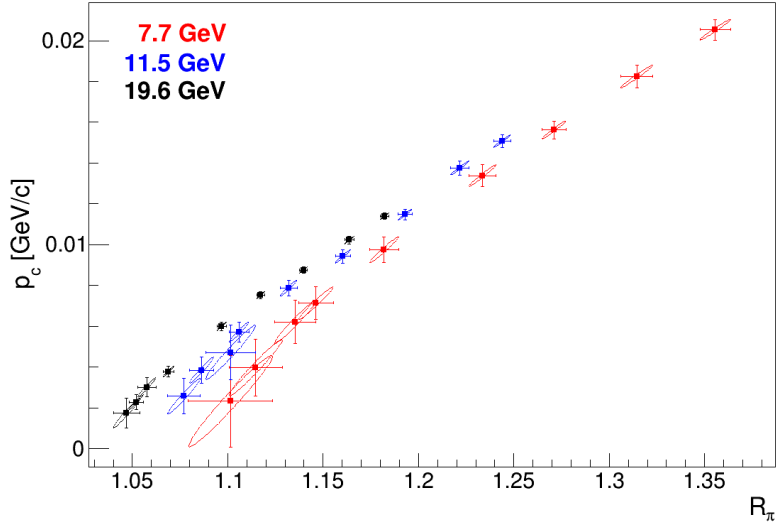

Figure 4: Contours in the $p_c$–$R_\pi$ plane showing 1-σ deviation lines from the minimum $\chi^2$ values, as well as the values of the fit parameters $p_c$ and $R_\pi$ (corresponding to the minimum $\chi^2$) with error bars.

To analyze the correlation between the two fit parameters, Figure 4 shows 1-σ uncertainty contours in the $p_c - R_\pi$ plane for Au+Au collisions at $\sqrt{s_{NN}} = 7.7$ GeV (red curves), $\sqrt{s_{NN}} = 11.5$ GeV (blue) and $\sqrt{s_{NN}} = 19.6$ GeV (black). There is a clear beam energy dependence in the values of the parameters. The contour lines do not overlap for the studied energies, however, in peripheral collisions there is a wider range of possible values compared to more central collisions. As the beam energy increases, the values of $p_c$ and $R_\pi$ are much more constrained than at 7.7 GeV. The $\chi^2$ contours indicate that the two parameters are correlated, if the Coulomb kick increases, the total pion ratio also increases.

Figure 5 shows the two fit parameters, $p_c$ and $R_\pi$, as a function of collision centrality expressed as $N_{part}$ in Au+Au collisions at 7.7 GeV, 11.5 GeV and 19.6 GeV. The centrality dependence of Coulomb kick is observed at all energies and it increases from peripheral to central collisions. The Coulomb kick values are greater at 7.7 GeV for all centralities than at higher BES energies. This may be due to large baryon stopping at midrapidity at the lower energy of 7.7 GeV. The decrease of Coulomb kick in more peripheral collisions can be also connected to the baryon stopping, which is centrality dependent, i.e. higher in more central collisions. At higher energies, the system expands with a stronger collective flow and charged pions are less influenced by the Coulomb field of the net-protons. In peripheral collisions, the collective flow, particle production and the net-proton densities are much reduced, resulting in smaller Coulomb kick values.

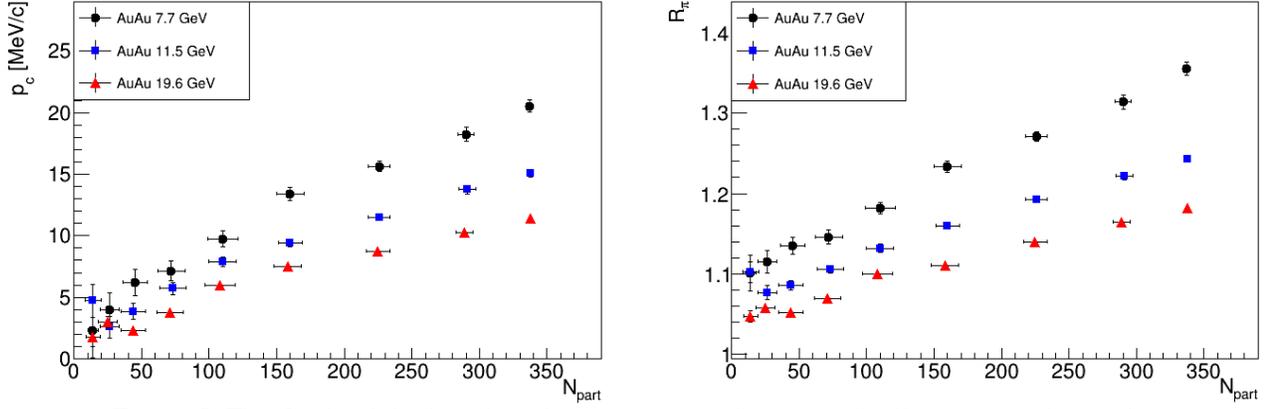

Figure 5: The Coulomb kick and total pion ratio as a function of collision centrality and energy.

The values of the pion ratios, $R_\pi$, decrease with increasing energy. At lower beam energies the ratios are larger than unity, which is likely due to isospin and significant contributions from resonance decays (such as $\Delta$ baryons). As the energy is increasing, there is a change in pion production mechanisms and direct pion pair production becomes important [17].

Because in our analysis, $dN^{ch}/dy$ is the net-charge of the interaction region, we subtract the antiproton dN/dy from proton dN/dy to get the net-proton values. The experimental values of proton and antiproton rapidity densities were taken from Ref. [17]. These net-proton dN/dy values were used to obtain the kinetic freeze-out radius based on Eq. 1. The energy and centrality dependence of the kinetic freeze-out radius (full symbols) is shown in Figure 6. The empty symbols represent the chemical freeze-out radius of the system obtained by STAR collaboration using a thermal model analysis [17]. The kinetic freeze-out radius shows no energy dependence for the energy range studied in this analysis. The same behavior was observed by STAR collaboration for the chemical freeze-out radius. The kinetic freeze-out radius decreases from central to peripheral collisions indicating that in a central collision a larger system is formed, which expands and decouples later in time and therefore, the pions have more time to feel the Coulomb interaction from the reaction partners.

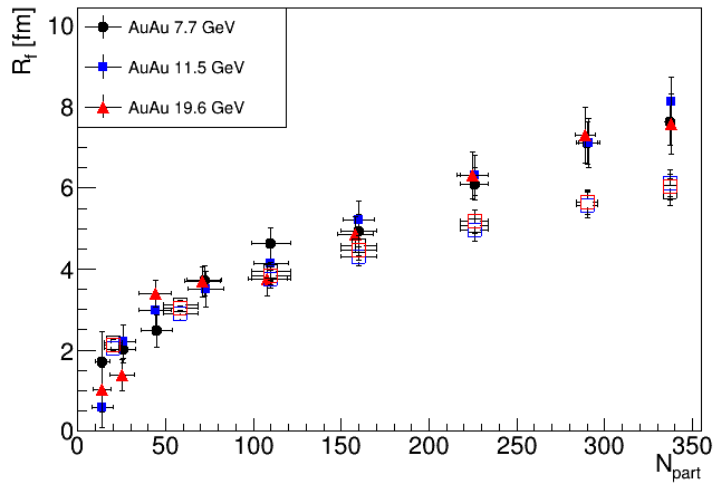

Figure 6: The kinetic freeze-out radius (full symbols) as a function of collision centrality for Au+Au collisions at 7.7 GeV (black symbols), 11.5 GeV (blue) and 19.6 GeV (red symbols). The empty symbols are the chemical freeze-out radii obtained by STAR experiment [17].

For all energies, the kinetic freeze-out radius is larger than the chemical freeze-out radius in central collisions. The separation between the chemical freeze-out radius and kinetic freeze-out radius increases in more central collisions, suggesting that the system expansion from chemical freeze-out stage to thermal freeze-out is stronger in these collisions.

The collision energy dependence of the Coulomb kick is shown in Figure 7. The black points are from most central BES Au+Au collisions, 0-5% centrality (this analysis) and the AGS (Au+Au collisions at 11.6 AGeV) and SPS (Pb+Pb collisions at 158 AGeV) results were taken from Ref. [14]. To study the energy dependence of $p_c$, three functions ($f_1$ – red curve, $f_2$ – blue dotted curve, $f_3$ – black dotted curve) were plotted also on Fig. 7. The curves are described by the functions:

$$f_1 = a + b \ln \sqrt{s_{NN}} \quad (3)$$
$$f_2 = \exp(a + b \ln \sqrt{s_{NN}}) \quad (4)$$
$$f_3 = a(\sqrt{s_{NN}})^b \quad (5)$$

where the value of the *a* and *b* parameters, as well as the $\chi^2/dof$ are given in Table 1:

| Function | a | b | $\chi^2/dof$ |
|---|---|---|---|
| $f_1$ | 36.96 ± 1.35 | -8.64 ± 0.49 | 12.16 |
| $f_2$ | 4.24 ± 0.08 | -0.61 ± 0.03 | 4.74 |
| $f_3$ | 69.15 ± 5.83 | -0.61 ± 0.03 | 4.74 |

Table 1: The parameters of the functions used to describe the energy dependence of the Coulomb kick.

The $f_2$ and $f_3$ functions seems to describe better the behavior of the Coulomb kick energy dependence, while the $f_1$ functions fails to describe this dependence.

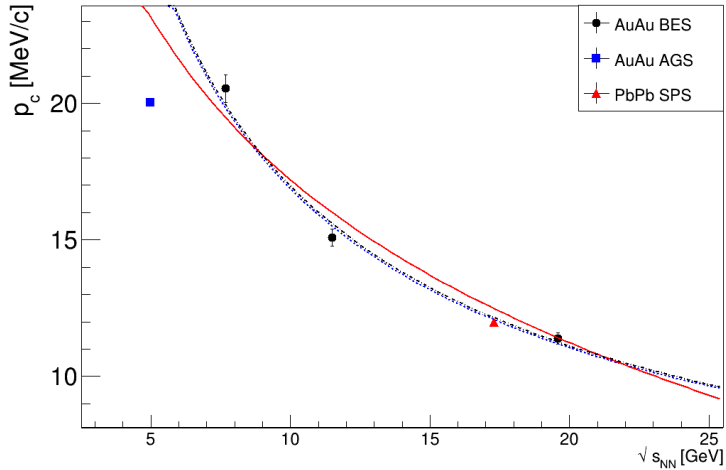

Figure 7: The Coulomb kick as a function of beam energy. The dashed and solid lines are the fits with two functions described in the text.

The Coulomb kick decrease with the increase of the beam energy, showing that the Coulomb interaction is stronger at lower energies, as it has been already observed in other analyzes [18-20]. If the colliding nuclei are strongly stopped, the total charge stays for a sufficient time to affect the charged pions.

## 3. Conclusions

The results indicate stronger Coulomb effects on the $\pi^-/\pi^+$ ratio in heavy systems at lower beam energies. For the same collision energy, the Coulomb interaction is larger in central collisions because there is strong stopping and an important positive net-charge in the central rapidity region. The Coulomb interaction decreases in peripheral collisions. The kinetic freeze-out radius is not changing with energy for the energy interval considered in this analysis and shows an increase from peripheral to central collisions indicating the formation of a larger system in more central collisions.

**Acknowledgements.**
This work was supported by IFA, RO-FAIR program, project number 09FAIR/16.09.2016.

**Conflicts of Interest**
The authors declare that there is no conflict of interest regarding the publication of this paper